 \documentclass[%
 mathleft,%
 final,%
 ]{an}
 \usepackage{graphicx}
\usepackage{times}
\newcommand{\kms}{km\,s$^{-1}$}

\begin{document}

% The following seven commands are intended for editorial usage and
% should be ignored by the author(s).
 \Pagespan{1}{}% Document's page range.
% % If second parameter is left empty, the last page is computed
% % automatically.
 \Yearpublication{2013}%
 \Yearsubmission{2012}%
 \Month{1}%
 \Volume{334}%
 \Issue{1}%
 \DOI{This.is/not.aDOI}%

\title{Accretion, winds and outflows in young stars}

\author{H. M. G\"unther\inst{1}\fnmsep\thanks{  \email{hguenther@cfa.harvard.edu}
}}
\titlerunning{Accretion, winds and outflows in young stars}
\authorrunning{H. M. G\"unther}
\institute{
Harvard-Smithsonian Center for Astrophysics, 60 Garden St., Cambridge, MA 02138, USA}

\received{XXXX}
\accepted{XXXX}
\publonline{XXXX}

\keywords{stars: formation, stars: pre-main sequence, stars: winds, outflows, T Tauri stars}

\abstract{Young stars and planetary systems form in molecular clouds. After the initial radial infall an accretion disk develops. For classical T Tauri stars (CTTS, F-K type precursors) the accretion disk does not reach down to the central star, but it is truncated near the co-rotation radius by the stellar magnetic field. The inner edge of the disk is ionized by the stellar radiation, so that the accretion stream is funneled along the magnetic field lines. On the stellar surface an accretion shock develops, which is observed over a wide wavelength range as X-ray emission, UV excess, optical veiling and optical and IR emission lines. Some of the accretion tracers, e.g. H$\alpha$, can be calibrated to measure the accretion rate. This accretion process is variable on time scales of hours to years due to changing accretion rates, stellar rotation and reconfiguration of the magnetic field. Furthermore, many (if not all) accreting systems also drive strong outflows which are ultimately powered by accretion. However, the exact driving mechanism is still unclear. Several components could contribute to the outflows: slow, wide-angle disk winds, X-winds launched close to the inner disk rim, and thermally driven stellar winds. In any case, the outflows contain material of very different temperatures and speeds. The disk wind is cool and can have a molecular component with just a few tens of \kms, while the central component of the outflow can reach a few 100~\kms. In some cases the inner part of the outflow is collimated to a small-angle jet. These jets have an onion-like structure, where the inner components are consecutively hotter and faster. The jets can contain working surfaces, which show up as Herbig-Haro knots. Accretion and outflows in the CTTS phase do not only determine stellar parameters like the rotation rate on the main-sequence, they also can have a profound impact on the environment of young stars. This review concentrates on CTTS in near-by star forming regions where observations of high spatial and spectral resolution are available.}

\maketitle

%TOD: kick out BDs? Not that much known anyway.

\section{Introduction}
While our Sun and our Earth are nearly five billion years old, star and planet formation is still ongoing in other regions of the galaxy. This allows us to directly observe the processes that shape the formation of stars and planetary systems. In addition, we know now several hundred extra-solar planets, so it is increasingly important to understand what controls the development of young stars and their proto-planetary disks. The disks and stars are connected in several ways. The most obvious one is stellar irradiation which critically influences the disk structure, but the disk structure also provides feedback on the star: If the inner disk is cleared, then the accretion will cease. This article reviews young cool stars (spectral type F-K), in one of their last stages of pre-main sequence stellar evolution, the so-called classical T Tauri star (CTTS) phase. Stars of type A and B in a comparable evolutionary stage are called HerbigAe/Be stars, for objects of lower mass I refer to the summary of the splinter session that was dedicated to brown dwarfs in the same volume.

After the initial collaps of a molecular cloud proto-stars develop an accretion disk. Material from the cloud falls onto that disk, where presumably planet formation takes place. The inner edge of the disk is ionized at a few stellar radii by the high-energy radiation of the proto-star. CTTS have strong magnetic fields, which can couple to the inner disk around the co-rotation radius. Mass can then flow along the field lines and impact on the star close to free-fall velocities. On the surface a strong shock forms which releases the energy. %This spot can be observed at several wavelength. 
This magnetically funneled accretion scenario is discussed in more detail in Sect.~\ref{sect:accretion}.

The proto-star irradiates not only the inner disk edge but also the disk surface. This can drive a disk wind. In general, outflows are required to remove angular momentum from the system, otherwise there would be no accretion. Winds may also be launched magnetically from the interaction region of a disk and a stellar magnetic field (X-wind) or directly from the star. Section~\ref{sect:outflows} summarizes common outflow tracers and their properties.

Last, some part of the wind can be collimated into jets with very small opening angles. These jets have an onion-like structure where the innermost component is the fastest part of the jet. In CTTS they are typically seen for a few hundred AU. Jets are the topic of Sect.~\ref{sect:jets}.

The review ends with a short summary of the open questions which I expect to drive the field in the next few years (Sect.~\ref{sect:questions}).

\section{Accretion}
\label{sect:accretion}
The simplest version of the magnetically funneled accretion model assumes a star with a dipole field that is aligned with its rotation axis. The field couples to the disk around the co-rotation-radius (Uchida 1983; Koenigl 1991). If the foot point is located outside the co-rotation radius the star looses angular momentum in this way; if it is inside, the star gains angular momentum. Observational indications that star and disk are magnetically coupled come from observations of X-ray flares with very large flare loops (Favata et al. 2005; see also the sketch in Fig.~\ref{fig:sketch}). In quiescence the connecting magnetic field lines carry the accreting mass and funnel it to impact points at high stellar latitude (Fig.~\ref{fig:sketch}). While the stellar magnetic field is certainly more complex, multipole components of higher order are important on smaller scales only. Magneto-hydrodynamic (MHD) simulations of inclined dipoles by Romanova et al. (2004) and later more complex fields have confirmed the theoretical idea of magnetically funneled infall (Long, Romanova \& Lovelace 2007; Long, Romanova \& Lamb 2012), but also revealed new accretion modes, where the accreting matter pushes the magnetic field lines apart and accretes in the plane of the disk (Kulkarni \& Romanova 2008). 

Where the accretion funnel hits the star the material passes through a strong accretion shock and heats up to temperatures of 2-3~MK, hot enough to emit X-rays. In the post-shock cooling zone the material cools radiatively until it mixes with the photosphere. A significant fraction of the emission is absorbed in the photosphere which in turn emits a continuum emission, that veils photospheric absorption lines of the underlying star. Because the accretion is magnetically funneled and everything moves along the field lines, a 1-D geometry should be a good approximation for the post-shock cooling zone. Several groups developed models along these lines with emphasis on the UV and optical continuum (Calvet \& Gullbring 1998), the X-ray emission (Lamzin 1998), high-resolution X-ray spectra and non-equilibrium ionization (G\"unther et al. 2007) and time-variability (Sacco et al. 2008).

The most direct tracer of the post-shock plasma can be found in X-ray data.
The first high-resolution X-ray observation of a CTTS (Kastner et al. 2002) showed a striking difference to main-sequence X-ray spectra in the He-like ions, in particular O\,{\sc vii} and Ne\,{\sc ix}. These ions emit triplets of collisionally excited lines and consist of a resonance ($r$), an intercombination ($i$) and a forbidden ($f$) line. The $f/i$ ratio is sensitive to the ambient density and UV photon field. For higher densities the lifetime of the upper level of the $f$ line is longer than the time between collisions. Thus, some of the electrons in the upper level of the $f$ line get excited into the upper level of the $i$ line and this decreases the $f/i$ ratio. UV photons can cause the same effect, so that an observation of a large $f/i$ ratio requires that the emitting region has both a low density \emph{and} a low UV-field. In CTTS we can directly measure the UV field, which turns out to be insufficient to influence the $f/i$ ratio. Thus, the small $f/i$ ratios observed in almost all CTTS (see Table 2 in G\"udel \& Naz\'e 2009) must be caused by high densities.

% \begin{table}
% \caption{\label{tab:CTTS}}
% \begin{tabular}{lll}
% \hline
% Star & observatory & reference \\
% \hline
% \multicolumn{3}{c}{low $f/i$ ratios}\\
% \hline
% TW Hya & Chandra/HETGS & \cite{2002ApJ...567..434K}\\
%        & XMM/RGS & \cite{twhya}\\
%        & Chandra/LETGS & \protect{\cite{2009A&A...505..755R}}\\
%        & Chandra/HETGS & \cite{2010ApJ...710.1835B}\\
% BP Tau & XMM/RGS & \cite{bptau}\\
% RU Lup & XMM/RGS & \cite{RULup}\\
% V4046 Sgr & Chandra/HETGS & \cite{v4046}\\
%           & XMM/RGS & \cite{2012ApJ...752..100A}\\
% MP Mus & XMM/RGS & \protect{\cite{2007A&A...465L...5A}}\\
% Hen 3-600 &  Chandra/HETGS & \cite{2007ApJ...671..592H}\\
% IM Lup & Chandra/HETGS & \protect{\cite{2010A&A...519A..97G}}\\
% V2129 Oph & Chandra/HETGS & \protect{\cite{2011A&A...530A...1A}}\\
% \hline
% \multicolumn{3}{c}{coronal $f/i$ ratios}\\
% \hline
% T Tau & XMM/RGS & \cite{guedel06}\\
% \hline
% \end{tabular}
% \end{table}

It is unclear how deep the accretion shock is buried in the atmosphere and how much, of the X-ray emission can escape (Drake 2005). While the observed high-densities in non-flare X-ray spectra are only seen in the CTTS, they do not have to originate below the accretion shock. Brickhouse et al. (2010) observed the CTTS TW~Hya in a deep X-ray grating spectrum and found that the densities in Mg~{\sc xii} are actually higher than in Ne~{\sc ix}, which in turn are higher than in O~{\sc vii} while post-shock cooling models predict exactly the opposite: Mg~{\sc xii} should be formed at the highest temperatures just behind the accretion shock, while O~{\sc vii} is predicted for the cooler and denser layers. Consequently, Brickhouse et al. (2010) suggest that part of the post-shock gas overcomes the magnetic confinement and leaks out into higher layers (Fig.~\ref{fig:sketch}). At the same time, Orlando et al. (2010) observe this effect in 2-D MHD simulations for weak magnetic fields.

The accretion funnels themselves are best observed in the optical. T~Tauri stars were initially defined as irregularly variable stars with the Balmer lines and the Ca~{\sc ii} H and K lines in emission (Joy 1945, 1949). Later, observers distinguished the CTTS from the weak-line TTS (WTTS) by means of their H$\alpha$ equivalent width with a boundary of 10~\AA{} between CTTS and WTTS. All these phenomena are closely related to the accretion process. Surveys of CTTS always show wide and asymmetric H$\alpha$ lines with wings reaching out to 400-500~km~s$^{-1}$, often with absorption components, e.g. Walker (1972), Appenzeller \& Wolf (1977) or Wichmann et al. (1999). Radiative transfer models for the magnetic funnels can explain the observed H$\alpha$ profiles (Muzerolle, Calvet \& Hartmann 1998). However, Bary et al. (2008) surveyed CTTS in the IR, and they calculate the Balmer and Paschen decrements and estimate the temperature and density in the emission region. They find high densities, which translate to cooling times of the order of a few minutes, while the free-fall time from the inner disk rim is a few hours. Thus, emission from the accretion funnel might not be the dominant contributor to the observed line profiles. Similar to H$\alpha$, observations of the He~{\sc i} 10830~\AA{} line show very wide lines often with red-shifted sub-continuum absorption (Edwards et al. 2006; Fischer et al. 2008; Ingleby et al. 2011). In a few cases hydrogen lines have been observed with spectro-astrometry. For this technique a spectrum is observed and the centroid of the spectrum on the CCD is measured for different wavelengths. Since the centroid can be measured to much better than one pixel and the continuum around the line in question provides a good comparison, offsets as small as a few stellar radii can be found. Whelan, Ray \& Davis (2004) use this method on a sample of CTTS and find that significant fraction of the Br$\gamma$ emission is related to outflows. Goto et al. (2012) observed TW~Hya, the closest CTTS. TW~Hya has a low accretion rate and no jet. Here, offset and line profile can be best explained with a combination of emission from the inner disk rim and the accretion funnels. 

Other characteristics of CTTS are due to the accretion shock on the stellar surface. If the star is not seen pole-on then the stellar rotation will make the hot and bright accretion shocks appear or disappear and thus cause large photometric variations. This has been observed in many different programs, one example is the ROTOR survey which collected photometric data for more than two decades (Grankin et al. 2007). Also, Doppler imaging can show the distribution of spots on the stellar surface (Strassmeier et al. 2004). In several papers Donati and collaborators use the Ca\,{\sc ii} infra-red triplet which is formed in the heated photosphere close to the accretion spots to study the distribution of the accretion funnels on the stellar surface (e.g. most recently on V2129 Oph: Donati et al. 2011). In addition, they performed Zeeman Doppler imaging and thus can extrapolate the magnetic field geometry (see e.g. the image of DG~Tau in Donati et al. 2008). In a multi-wavelength observation of V2129~Oph Argiroffi et al. (2011) combine this information with time-resolved X-ray spectroscopy and find that the soft X-ray emission from the accretion spot is only observed when the accretion column is seen from the side. Thus, at least in this star, it seems that a significant amount of the X-ray emission is hidden when looking along the accretion funnel. This model explains both X-ray and optical lightcurves.

% Doppler Imaging
% AA Tau und neue Argiroffi as an example at the end of this section.
%Fischer He I red abs 2008ApJ...687.1117F

\section{Outflows}
\label{sect:outflows}
Outflows and also highly collimated jets (see next section) seem to be a natural consequence of disk accretion. To allow accretion to proceed in the first place, there must be a mechanism that removes angular momentum from the system. Some effects in the disk, e.g. turbulence, have been suggested to move angular momentum outwards, but this is not a topic of the current review. Alternatively, the angular momentum can be ejected by outflows. 

\begin{figure}
\includegraphics[angle=0, width=\linewidth]{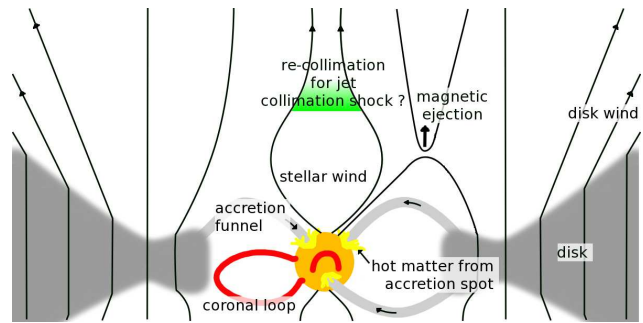}
\caption{Sketch of the different infall and outflow patterns (arrows) in CTTS (not to scale).  The magnetic field (black lines; lines with arrow heads for mass flux along field lines) can reconnect on different scales and flares large enough to link to the disk have been observed. The configuration of the magnetic field might not always be symmetric. The repective contribution of stellar wind, magnetic ejections and disk wind is unclear; in this sketch I speculate that the innermost jet component might be launched from the star and could be heated in a recollimation shock (see text for discussion).}
\label{fig:sketch}
\end{figure}

Three different launching regions have been proposed (see Fig.~\ref{fig:sketch} for a sketch): First, there could be a stellar wind in analogy to the solar wind (Kwan \& Tademaru 1988; Matt \& Pudritz 2005). The accretion shock provides additional heating in the upper layers of the atmosphere, so that CTTS can potentially accelerate stronger stellar winds than MS stars (Cranmer 2009). However, there is a fundamental limit. In hot, optically thin plasma with density $n$ the cooling increases as $n^2$. Higher mass loss rates require higher densities in the corona. At some threshold density the cooling rate becomes so high, that the temperature drops below the point where it can drive a wind. Matt \& Pudritz (2007) calculate an upper boundary to the mass loss by a hot stellar wind of $10^{-11} M_{\odot}$~yr$^{-1}$ based on this argument. Second, outflows can be driven from the inner disk edge close to the co-rotation radius  along the interface of the stellar magnetic field and the disk field. These winds are called X-winds (Shu et al. 1994). Third, winds can be driven from the disk (Blandford \& Payne 1982; Pudritz \& Norman 1983; Anderson et al. 2005). If the outer layer of the disk is warm enough to ionize some gas, it can be loaded onto the field lines of the disk field. Since these are stretching outwards, material travelling along the field lines is accelerated magneto-centrifugally. 

Emission from winds is typically faint due to their low density.
Winds are usually observed as blue-shifted absorption of either the continuum or the broad emission lines from the accretion shock. While this approach reveals both the velocity and the ionization stage of the wind, which is closely related to the temperature, it only probes wind components in the line-of-sight between us and the star. There is overwhelming evidence of cool winds from P-Cygni profiles in optical lines, most notably H$\alpha$. Alencar \& Basri (2000) find absorption components in 80\% of their CTTS sample with velocities in the range $0-270$~km~s$^{-1}$. While the absorbtion is typically strong in H$\alpha$ and H$\beta$, it falls off for higher members of the Balmer series indicating that the wind is optically thin at those transitions. Absorption is also seen in the Na\,{\sc i} D line or Ca\,{\sc ii} H and K. Ardila et al. (2002) surveyed eight stars in the NUV, where the Mg\,{\sc ii} H and K lines are located and also found strong absorption components out to 300~km~s$^{-1}$; in two cases absorbtion was detected at multiple velocities. The absorption is stronger in systems with low inclination. Another important tracer of outflows is the He~{\sc i} 10830\AA{} line which has a very high opacity and a meta-stable lower level that is long-lived if the collisional de-excitation rate, i.e. the density, is low. This level is 21~eV above the ground state, so He~{\sc i} 10830\AA{} absorption indicates a wind with a temperature around $20,000$~K. This diagnostic was pioneered by Edwards et al. (2003, 2006). These studies find a roughly comparable number of CTTS with narrow (as expected from a disk wind) or wide (as expected from a stellar wind) blue-shifted absorption indicating that both types may contribute to the mass loss and the viewing geometry and time variability determine which type is observed at any one moment. In simulations Kurosava, Romanova \& Harries (2011) present the first line profiles computed from multi-dimensional MHD models that can be matched to those observations.

Going up to higher energies, Dupree et al. (2005) suggested that the O\,{\sc vi} line at 1032~\AA{} shows the presence of a hot wind. This line is very asymmetric. While there is little or no emission on the blue side of the line, it is a few hundred km~s$^{-1}$ wide on the red side. They suggested that the blue side of an intrinsically wide, Gaussian line is absorbed by a hot wind. However, Johns-Krull \& Herczeg (2007) analyze observations of the C\,{\sc iv} doublet around 1550~\AA{}. If the blue absorption hypotheses is true, then one line in the doublet should be absorbed by the other when in fact they have very similar profiles. %Also, a wind would lead to a P-Cygni profile with continuum absorption. While the continuum due to the accretion shock in TW~Hya is faint around O\,{\sc vi} this can be tested at 1550~\AA{} and no continuum absorption is found. Thus, the  C\,{\sc iv} lines show that there is no hot wind in TW~Hya and the fact that the line profiles of C\,{\sc iv} and O\,{\sc vi} are so similar makes it unlikely that this is caused by temporal variability. 
However, some CTTS show blue-shifted \emph{emission} in O\,{\sc vi} thus gas in the temperature range $2\times10^5$~K must be present, possibly due to shocks in the outflows (G\"unther \& Schmitt 2008).

These uncertainties about the temperature of the outflows make it difficult to find an ideal tracer of the total mass loss rate, but the most important contributor is probably the optically visible outflow. Studies of this find that the outflow rate typically is few percent of the mass accretion rate (Cabrit et al. 1990; Hartigan, Edwards \& Ghandour 1995; Coffey, Bacciotti \& Podio 2008).

\section{Jets}
\label{sect:jets}
The innermost part of the wind can be a very fast and highly collimated jet with opening angles of a few degrees in some, but not all CTTS. It is possible that jets are present in all CTTS and the non-detection particularly in older sources with lower mass loss rates or in CTTS with a face-on geometry are just due to their faintness (Sauty et al. 2011). The relation between mass accretion rate and wind loss rate explains why very young proto-stars, which are stronger accretors, drive more powerful jets which reach out to a few parsecs in some cases. Their bow-shocks interact with the molecular cloud on a large scale (see review by Bally, Reipurth \& Davis 2007). 
This can inject significant turbulence in the surrounding molecular could. Here, we concentrate on CTTS, which drive jets on smaller scales, sometimes called ``micro jets''. The most prominent example is the jet from DG Tau. The structure resembles an onion where the inner layers are consecutively faster (Bacciotti et al. 2000) and can reach velocities up to a few hundred km~s$^{-1}$. The jet components also have vastly different temperatures. The outer layers are typically traced in H$\alpha$ and forbidden emission lines (FELs) like [O\,{\sc i}] or [S\,{\sc ii}]. The ratio of two [S\,{\sc ii}] lines is density sensitive; other line ratios trace the ionization fraction and the temperature (Bacciotti \& Eisl\"offel 1999).
The densities derived in DG Tau's jet are up to $10^5$~cm$^{-3}$ and the temperatures in the FELs can be explained by shock heating with shock speeds around 100~km~s$^{-1}$ (Lavalley-Fouquet, Cabrit \& Dougados 2000).  De Colle, del Burgo \& Raga (2010) show that simple methods which take the observed surface luminosity at face value generally underestimate the density, temperature and ionization in the innermost layers because the line-of-sight through the jet passes through different layers, so that we derive quantities averaged over the jet. In their tomographic reconstruction they assume axisymetry for the HH~30 jet and find a more centrally condensed and more structured jet than previous studies.

The inner layers of the jet are much hotter.
In DG~Tau's jet X-ray emission from the innermost layer has been detected by G\"udel et al. (2005, 2008, 2011).
There are three different components to the X-ray emission: First, there is a knot in the jet at 4-5\arcsec{} resolved by \emph{Chandra}. Second, there is a hard inner component, which shows flaring in the lightcurve and thus is identified with the central star. The high absorbing column density prevents the detection of soft X-rays from the star, and indeed the centroid of the soft X-ray emission is offset from the central star by 0.2\arcsec{}, i.e. 40~AU deprojected (Schneider \& Schmitt 2008). This third component can be explained by shock heating and indeed the dimensions of the post-shock cooling zone would be so small that it remains unresolved even in HST images (G\"unther, Matt \& Li 2008). Only 0.1\% of the mass flux is required to power the observed X-ray luminosity. RY~Aur is the only other CTTS where X-ray emission from the jet is seen (Skinner, Audard \& G\"udel 2011), but more detections exist for younger objects with higher mass loss rates.

The knots in the jet, observed as Herbig-Haro objects, are internal working surfaces. They can be caused by material ejected at different speeds. When a faster blob catches up with a slower one a shock front develops that thermalizes some of the kinetic energy. As side effects, a reverse shock will form that travels upstream and oblique shocks might interact with the cloud material at the boundaries of the cocon that separates the jet material from the cloud. Hydrodynamic simulations show these effects well and explain much of the observed morphology in different bands (Bonito et al. 2010a, 2010b). Alternatively, the jet could be precessing. If the angle is wide enough, then every knot plows through the cloud material and behaves like a bow shock. Again, this can nicely explain the morphology of some of the observed jet features (Raga et al. 2001).

There is more controversy about the emission components closest to the star. They could just be knots in the making or they could be stationary components that are related to the jet launching and collimation process. It it unclear how exactly the jet material is accelerated and collimated but most models rely on magnetic fields of some kind (Ferreira \& Petrucci 2011). It has been shown analytically and in (magneto-) hydrodynamical simulations that jets can be driven entirely from a disk wind (e.g. Ferreira 1997 or Ramsey \& Clarke 2011) or from the inner edge of the disk  (Lii, Romanova \& Lovelace 2011), but if stellar winds contribute to the mass loss at all, they are most likely responsible for the fastest and innermost outflow components as it is hard to imagine how to collimate a disk wind while at the same time pushing the stellar wind to larger opening angles. Once the mass is launched, it needs to be collimated which can happen through helical magnetic field lines frozen into the rotating outflow (e.g. Lovelace, Wang \& Sulkanen 1987), an external magnetic field which threads the disk (e.g. Fendt 2009) or simply external pressure. 

While several techniques provide a sufficient spatial resolution along the jet axis, even space-based observations offer at best marginal velocity information perpendicular to the flow direction. If seen, these velocity differences indicate a jet rotation and thus provide a handle on the angular momentum lost in CTTS jets (Bacciotti et al. 2002; Coffey et al. 2004, 2007; Woitas et al. 2005). The results are promising and indicate a jet rotation that is consistent with MHD disk wind launching models. In CTTS with strong bipolar outflows the rotation of both lopes appears consistent. However, the signals are weak in all cases and for HH~30 the (marginal) jet rotation is contrary to the measured disk rotation, highlighting the difficulties in measuring these weak signals. It is worth noting that Matt \& Pudritz (2008) can show in their simulations that stellar winds of sufficient mass loss can provide enough torque to explain the slow rotation of CTTS. 

With these different theoretical ideas in mind, we need to turn to observations to decide which mechanism is actually realized in CTTS and we should find hallmarks of the initial launching and collimation in those observations that resolve the innermost arcsec down to the star. In this region FELs are good tracers, because due to their low critical densities they do not have a stellar contribution and thus are not effected by uncertainties in the stellar PSF. FELs often have multiple velocity components with a low velocity component $<100$~km~s$^{-1}$, a medium velocity component which is about twice as fast and a fast velocity component of a few hundred km~s$^{-1}$ (see Coffey, Bacciotti \& Podio 2008 for a survey of five CTTS with the \emph{HST}/STIS long slit). Lavalley et al. (1997) already suspected that the inner low-velocity component might be stationary in DG Tau and this is strengthend by the observations of Schneider et al. (2012), while the medium velocity plasma seems to travel outward and thus has to be associated with a new knot forming. In addition to the FELs Schneider et al. (2012) observe C\,{\sc iv} emission with a deprojected velocity of 200~km~s$^{-1}$ and thus establish yet another jet component with a velocity and mass loss rate between the slow, but massive FELs and the fast X-ray emitting component with a very small mass flux. The C\,{\sc iv} peaks at about 40~AU from the central star and again is consistent with heating in a jet collimation shock. Although some doubts remain, indications are mounting that we see a stationary component in different temperature tracers that can actually tell us about the conditions in the jet collimation regime. 

As mentioned in the previous section, Coffey et al. (2008) find that about 1-7\% of the accreted mass is ejected in outflows. While it is not possible to separete the mass flux in the wide angle wind form the highly collimated jet for most stars, those with the best data indicate that the jet actually dominates the mass and angular momentum loss.

\section{Open questions}
\label{sect:questions}
After reviewing the accretion and outflow processes in CTTS, I conclude with a few open questions which in my opinion have the potential to drive our understanding of CTTS forward in the next few years. Initial work has been done on those questions (some of which is discussed in the previous sections) but there is no ``big picture'' answer yet.

\begin{description}
\item{How does the accretion shock interact with the photosphere and the corona?} Models of a 1D accretion shock successfully described the initial observations, but new data and new MHD simulations show that this is not the full picture. How does accretion relate to activity? Are the accretion funnels the origin of the long flare loops occasionally found on CTTS or do they in fact inhibit coronal activity by filling magnetic loops with relatively cool material?
\item{How are accretion and outflow related?} Clearly, the outflows are powered by the accretion but which process or which configuration of the magnetic field side-tracks some of the accreted mass into an outflow and what sets the relation between accretion and outflow rate?
\item{How is the jet launched?} Really, here are two separate problems to consider. One is to find out which of the outflow components is collimated into the jet and the other one is how this collimation actually happens. This is foremost a question for theorists, because observations now and in the near future will not reach the required resolution.
\item{What about binary accretion?} Many CTTS are components of binaries or higher order systems in which accretion disks directly influence each other. In wide binaries each component can be treated separately and very close binaries can accrete from the same disk. How does this work for intermediate situations or more complex systems?
\end{description}

% ---------------- Fig 1
% \begin{figure}
% \includegraphics[angle=0, width=\linewidth]{empty}
% \caption{A one-colum figure. Give width in mm or cm instead of ``linewidth'' if desired.}
% \label{Flabel1}
% \end{figure}

\acknowledgements
HMG acknowledges support from grant GO-12199.01-A from the Space Telescope Science Institute, which is operated by the Association of Universities for Research in Astronomy, Inc., under NASA contract NAS5-26555.

The author thanks Sean Matt for advice on how to show the different winds and magnetic fields in the figure.

% Use this code if you wish to generate your bibliography with BibTeX;
% please replace first the string "an-demo" below with the name(s) of
% the BibTeX data base(s) you want to use.
% The resulting bibliography-output (the contents of the .bbl file)
% must be pasted into this file before submission.
%
%\bibliographystyle{an}
%\bibliographystyle{../../../my_articles/an_hack}
%\bibliography{../../../my_articles/articles}
%
% Replace the following example bibliography with your references
% before submission:

\end{document}